\documentstyle[12pt]{article}

\begin{document}

%begin macros
\def\nn{\nonumber \\}
\def\be{\begin{equation}}
\def\ee{\end{equation}}
\def\ba{\begin{eqnarray}}
\def\ea{\end{eqnarray}}
\def\la{\label}
\def\re{(\ref}

\def\i{{\rm i}}
\let\a=\alpha \let\b=\beta \let\g=\gamma \let\d=\delta
\let\e=\varepsilon \let\z=\zeta \let\h=\eta \let\th=\theta
\let\dh=\vartheta \let\k=\kappa \let\l=\lambda \let\m=\mu
\let\n=\nu \let\x=\xi \let\p=\pi \let\r=\rho \let\s=\sigma
\let\t=\tau \let\o=\omega \let\c=\chi \let\ps=\psi
\let\ph=\varphi \let\Ph=\phi \let\PH=\Phi \let\Ps=\Psi
\let\O=\Omega \let\S=\Sigma \let\P=\Pi \let\Th=\Theta
\let\L=\Lambda \let\G=\Gamma \let\D=\Delta

\def\0{\over } \def\1{\vec } \def\2{{1\over2}} \def\4{{1\over4}}
\def\5{\bar } %\def\5{\overline }
\def\6{\partial }
\def\7#1{{#1}\llap{/}}
\def\8#1{{\textstyle{#1}}} \def\9#1{{\bf {#1}}}

\def\({\left(} \def\){\right)} \def\<{\langle } \def\>{\rangle }
\def\[{\left[} \def\]{\right]} \def\lb{\left\{} \def\rb{\right\}}
\let\lra=\leftrightarrow \let\LRA=\Leftrightarrow
\let\Ra=\Rightarrow \let\ra=\rightarrow
\def\ul{\underline}

\let\ap=\approx \let\eq=\equiv  %% \let\ex=\times \let\hc=\dagger
        \let\ti=\tilde \let\bl=\biggl \let\br=\biggr
\let\bi=\choose \let\at=\atop \let\mat=\pmatrix
\def\CL{{\cal L}} \def\CD{{\cal D}} \def\rd{{\rm d}} \def\rD{{\rm D}}
\def\CH{{\cal H}} \def\CT{{\cal T}}
\newcommand{\dR}{\mbox{{\sf I \hspace{-0.8em} R}}}
%end macros

\begin{titlepage}
\renewcommand{\thefootnote}{\fnsymbol{footnote}}
\renewcommand{\baselinestretch}{1.3}
\hfill  TUW - 93 - 09\\
\medskip
\hfill hep-th/yymmddd\\
\medskip
\vfill

\begin{center}
{\bf {\LARGE{Quantization and the Issue of Time for Various Two-Dimensional
Models of  Gravity}}}
\medskip
\vfill

\renewcommand{\baselinestretch}{1} {\large {
THOMAS STROBL\footnote{e-mail:
tstrobl@email.tuwien.ac.at} \\ \medskip\medskip
\medskip \medskip
Institut f\"ur Theoretische Physik \\
Technische Universit\"at Wien\\
Wiedner Hauptstr. 8-10, A-1040 Vienna\\
Austria\\} }
\end{center}

\vfill
\begin{center}
Talk given at the Journ\'ees Relativistes '93, 5 - 7 April, Brussels, Belgium
\end{center}
\vfill
\renewcommand{\baselinestretch}{1}

\begin{abstract}
It is shown that the models of 2D Liouville Gravity, 2D Black Hole- and
$R^2$-Gravity are {\em embedded} in the Katanaev-Volovich model of
 2D NonEinsteinian Gravity.
Different approaches to the formulation of a quantum theory for
the above systems are then presented:
The Dirac constraints can be solved exactly in the momentum representation,
the path integral can be integrated out, and
the constraint algebra can be {\em explicitely} canonically
abelianized, thus allowing also for a (superficial) reduced phase space
quantization. Non--trivial dynamics are obtained by means of time dependent
gauges. All of these approaches lead to the {\em same} finite dimensional
quantum mechanical system.
\end{abstract}

\vfill
\hfill Vienna, August 1993  \\
\end{titlepage}

\renewcommand{\baselinestretch}{1}
\setcounter{footnote}{0}
\renewcommand{\thefootnote}{\alph{footnote}}

The Jackiw-Teitelboim model$^1$
\be \CL_{JT} = -e \, \pi_2  ({R\0 2} - \D)     \la{JT}  \ee
is a classical example for a two-dimensional playing laboratory
of four-dimensional quantum gravity. Reformulated as an
SO(2,1) $BF$  theory, it was quantized in the connection representation of
this group by means of Wilson lines.$^2$ The resulting Hilbert space
 was then given quite formally as spanned by the character
functions of all
(including infinite dimensional)
irreducible representations
of SO(2,1);  considerations about dynamics for this system, moreover, have
been missing up to now.
Beside \re{JT}), recently there has been  growing interest in the
string inspired models of 2D Black Hole (BH) gravity,$^3$ which can be
reformulated as
\be \CL_{BH} = -e({\pi_2 R \0 2}  - \l)     \la{BH}  \ee
or as a (centrally extended) ISO(1,1) ($BF$) gauge theory,
and $R^2$-gravity$^4$  \be \CL_{R^2} = -e({\g \0 4} R^2+\l). \la{R2}\ee
In \re{JT}) to \re{R2}) $R$ is the (torsion free) Ricci scalar, $\pi_2$
an auxiliary field, and $e = \det e_\mu{}^a  = \sqrt{-\det g_{\m\n}}$.

Less well-known than the above  Lagrangians for 2D gravity is
\be {\cal L} \, = - e \, (\frac{\gamma}{4} \, R^2 -
\frac{\beta}{2} \,T^2 + \lambda),   \la{L}  \ee
introduced by Katanaev and Volovich$^5$ as the
most general second order Lagrangian for the {\it zweibein} $e^a$ and
spin connection $\o$
(Ricci scalar
$R=2\ast d\o$,
torsion
$T^a =\ast De^a$).
As  toy model for 4D quantum gravity, however, it should be regarded as
at least on the same footing as \re{JT}) to \re{R2}). Whereas D.\ Schwarz
reported on a perturbative  treatment of
\re{L}) at the last meeting,$^6$ here we show\footnote{as the outcome
of a collaboration with P.\ Schaller$^7$}
that an {\em exact} quantization of $\CL$ is indeed possible and that many
conceptual problems of a quantum theory of gravity, like e. $\!\!$g.\ the
issue of
time, can be discussed very explicitely within the resulting framework.
Moreover, as is seen from the first order form of \re{L})
\be
\CL_H = e({\pi_2 \0 2} R + \pi_a T^a + E), \qquad
  E \; \equiv \; \frac{1}{4\gamma} \, (\pi_2)^2
- \frac{1}{2\b} \, \pi^2 - \lambda,   \la{LH}
\ee
the models \re{JT}) to \re{R2}) are imbedded in the Katanaev-Volovich (KV)
model as specific limits of the coupling constants! $R^2$-gravity
is obtained in the limit $\b \to \infty$ and $\CL_{BH}$ by further
letting $\g \to \infty$ [afterwards integrating out the auxiliary fields
$\pi_A \equiv (\pi_a,\pi_2)$ and $\pi_a$, respectively]. Shifting $\pi_2$ by
$-2\g \D$ in $\CL_H$, furthermore, the JT-model results from $(\b, \g) \to
\infty, \, \l-\g\D^2 \to 0$ (up to a  total
divergence).\footnote{This limit was noted also in a footnote in Ref.\ 8.
The limit $\b \to \infty$, on the other hand, was also considered
in Ref.\ 9 on the level of the Hamiltonian field equations; as presented here,
however, the limit {\em does} "commute with the procedure leading to the
equations of motion" and is clearly seen to correspond to \re{R2}).}
With some appropriate rescalings these
contractions  allow one to solve all the above four models {\em simultanously}
when solving \re{L}).

{}From \re{LH}) we learn that $\pi_A$ are the momenta conjugate to the
one-components of the {\it zweibein} and spin connection, whereas the
zero-components of the latter are Lagrange multipliers for the (first class)
constraints
\addtocounter{equation}{1}
$$ G_a \,=\,  \partial \pi_a +  \varepsilon_a{}^b  \, \pi_b \,\omega_1  +
 \varepsilon_{ab} \,  E \, e_1{}^b \; \approx \; 0 \eqno(\arabic{equation}a) $$
$$ G_2 \,=\, \partial  \pi_2 +
\varepsilon^a{}_b \, \pi_a \, e_1{}^b
\; \approx \; 0. \eqno(\arabic{equation}b) $$
Restricting our attention to the topology $S^1 \times \dR$, the Hamiltonian
reads $H = - \oint_{S^1} (e_0{}^a G_a + \o_0 G_2) dx^1$; it is a
combination of the constraints since the flow parameter $x^0$ is subject
to the diffeomorphism symmetry of the system. The difference between the
models \re{JT}) to \re{L}) enters only through the quantity $E$ at this
point, which is given by  the second equation \re{LH})
in the case of the KV-model.

The constraint algebra has structure functions in the case of \re{L}) and
\re{R2}), but becomes the expected SO(2,1) and ISO(1,1) algebra in the
JT- and BH-limit, respectively. Simple  redefinitions of the constraints,
however,  allow to abelianize the respective algebras: Integrating for the
quantity $Q$ in
$ \6 Q =\exp({\pi_2\0 \b})[\pi^a G_a - E G_2]  \approx 0 $
one obtains
\be Q = -\b \exp({\pi_2\0\b})[E - {\b\0 2\g}\pi_2 + {\b^2 \0 2\g}]. \la{Q} \ee
On patches of the phase space where $\pi^2 \neq 0$ one
can use then either of the sets of new {\em canonical} variables
\be ({\pm G_\pm\0\pi_\pm}, \mp {G_2\0\pi_\pm}, Q; \pi_2,\pi_\pm,P^{(\pm)}),
\qquad P^{(\pm)}\equiv - \exp(-{\pi_2\0 \b}) \, {e_1{}^\mp \over \pi_\pm}.
\la{can} \ee
The constraint algebra generated by the first two coordinate fields and
$\6 Q$ is abelian now. From this canonical
splitting we see also that the only 'physical' (i. $\!$e.\ gauge independent)
variables of the KV-model are the zero modes $Q_0$ and $P_0 := P_0^{+}
\approx P_0^{-}$. As a careful analysis of {\em all} the classical solutions
on the cylinder shows,$^{10}$ $P_0$ extends also to the
case where $\pi_a =0$ at some points of the space-time manifold, if one
takes the Cauchy principle value prescription at the poles of
integration. Thus excluding
only the deSitter solution $-\b T_a = \pi_a =  0, \g R=\pi_2 =
\pm \sqrt{4\g\l}$ (which at least in the momentum representation is a singular
point anyway), we can immediately quantize the system \re{L}):
The Hilbert space is an $\CL^2(\dR)$ spanned by the wave functions
$\psi(Q_0)$, square integrable with respect to the Lebesgue measure as
dictated (up to unitary equivalence) by the hermiticity requirement
of the Dirac
observable $P_0 = -i \hbar d/dQ_0$.

Rescaling the constant of integration in $Q$ by $-\b\l$ for $R^2$-
and ISO(1,1)-Gravity, and  $Q \to \exp(-\g\D)Q -\b\l + \b\g\D^2
-\b^2\D +\b^3/2\g$ for the SO(2,1)-Gravity, all of the above holds
also for the models defined by \re{JT}) to \re{R2}) in the corresponding
limits. The (only)\footnote{An exception to this is given for $\l = 0$ in
\re{BH}), which corresponds to $E\equiv 0$ in \re{LH}); we will exclude this
case in the following and give more details elsewhere.$^{10}$}
Dirac observables of these models are then, respectively,
\be Q_0^{R^2}\! =\! \oint {\pi^2\0 2}  + \pi_2 \l -{(\pi_2)^3\0 12\g},\,
Q_0^{BH} \!=\!   \oint {\pi^2\0 2}  + \pi_2 \l,\,
Q_0^{JT} \!=\!   \oint {\pi^2\0 2}  - {\D(\pi_2)^2\0 2}\!\equiv\! \2 \oint
\pi_A \pi^A,
\la{Qs} \ee
with the (in each case) conjugate momentum $P_0 = -\oint {e_1{}^-\0 \pi_+}$.

An alternative approach to find the Hilbert space of physical wave functions
is provided by the Dirac quantization. For this purpose we choose a momentum
representation for the wave functionals ($e_1{}^a \to i\hbar\d /\d \pi_a$,
etc.) and the operator ordering within
the quantum  constraints  as written in (6).  The constraint algebra
has no anomalies then so that it is consistent to search for their kernel.
Except for distributional
functionals located at $\pi_a =0$ it is given by the functionals
\be \Psi_{phys} = \exp(\mp {i\0 \hbar} \oint \ln \! \mid \!
\pi_\pm \!\mid \! d\pi_2) \;
\ti \Psi [Q] ,\qquad \6 Q  \, \ti \Psi[Q]=0.
\la{sol} \ee
The second equation is a restriction on the support of $\ti \Psi[Q]$,
due to which the first equation holds for either of the two signs with
the {\em same} $\ti \Psi[Q]$. Again a state vector is determined by the
specification of a function depending on $Q_0$.
Within $\Psi_{phys}$ the parameters $\b, \g,\l$ enter only through $Q$;
thus the above {\em explicite} form of the physical wave functions
holds also for the JT-, the BH-, and the $R^2$-models; one merely has to
replace $Q_0$ by the corresponding quantity in \re{Qs}).\footnote{A
difference on the quantum level appears (at least in the momentum
representation)  only in the distributional solutions located at $\pi_a=0$.
These we ignore here since they are  a one parameter family {\em only} in the
 case $(\b,\g) \to \infty, \l =0$, excluded in this note.} An inner product
can be constructed as above after the abelianization by the hermiticity
requirement of the Dirac observables as a restriction on the measure,
or, as might be advantageous in the more complicated 4D Gravity where
it is difficult to find all Dirac observables, by a gauge fixing
procedure similar to the one within a Lagrangian path integral.$^7$
%(c.\ f.\ Ref.\ 7 for more details on this point).

Since the Hamiltonian vanishes on \re{sol}), there is no meaningful
{\it Schr\"odinger} equation at this point (the so-called 'problem of time').
However, in a reparametrization invariant theory already {\em classically} a
time coordinate $x^0 \equiv t$ enters only after the specification of a
coordinate system, which is {\em equivalent} to the choice of a gauge. Thus
we choose e. $\!$g.
\be \pi_+=1, \quad \6 e_1{}^- =0, \quad \pi_2 = t,  \la{gauge} \ee
which provides a global foliation of the (classical) space-time manifold for a
good part of the solutions,$^{7,10}$ and express the remaining
phase space variables (by inverting the constraints) and the Lagrange
multipliers (by requiring that the $t$-dependence of \re{gauge}) is
generated by $H$) in terms of the Dirac observables $Q_0, P_0$ and the
'gauge fixing parameter' $t$. In this way we obtain e. $\!$g.  $e_1{}^-=
-\exp(t/\b)P_0$ as a now well-defined operator on our $\CL^2(\dR)$, being
{\em parametrized} by $x^0$ (in other gauges also $x^1$). A further
analysis shows that the time evolution corresponding to \re{gauge}) is
indeed {\em unitary} and generated by a (time dependent) effective Hamiltonian
$h(t)$. For more details the reader is referred to Ref.\ 7.

The Hamiltonian path integral can  also be integrated out straightforwardly.
In the
(time dependent)
gauge \re{gauge}) one obtains in agreement with
above [$\ph^A \equiv (e_1{}^a,\o_1)$]
\ba           \langle \pi_{A,2},t_2;\pi_{A,1},t_1 \rangle & =&
      \!\!\!\!\!\!\int\limits_{\pi_A(t_1) = \pi_{A,1}}^{\pi_A(t_2) = \pi_{A,2}}
   \!\!\!\!\!\!\!\! [D\pi_A][D\ph^A] \exp{(-i\! \int\!\! d^2x \, \ph^A \dot
   \pi_A)} \d[G_A]\d[\re{gauge})] (\mbox{FP-det})  \nn  &=&\int
   [dQ_0][dP_0] \exp{[-i\! \int\!\! dt(Q_0 \dot P_0 + h(t))]}.
\la{pi}
\ea
In a Hamiltonian path integral one can also have non-trivial dynamics
with time independent gauges, if one either
adds an appropriate surface term
to the Lagrangian or if one
gauge fixes the Lagrange multipliers instead of the constraints; in
the latter approach an invariant integration measure has to be constructed
by hand and the  gauge equivalence of boundary conditions $\pi_{A,i}$ analyzed
seperately.$^7$

Heuristically speaking, different choices of time should correspond to
different observers. A careful comparison of the resulting
quantum theories would be an interesting task and the above models
a possible area for such an investigation.

The reader further interested
in the Katanaev-Volovich model is referred to Ref.\ 11.
A complete description of the reduced phase spaces
of \re{JT}) and  \re{L}) with cylindrical space-time topology will be given
in Ref.\ 10.

\vskip.4cm
The author  thanks  P. Schaller
for his  collaboration and T.\ Kl\"osch for collaboration on related work.

\vspace{- 10 pt}
\renewcommand{\Large}{\normalsize}

\end{document}